\documentclass[journal]{IEEEtran}
\usepackage{graphicx}
\usepackage{amsfonts}
\usepackage{amsmath,amsthm}
\usepackage{enumerate}
\usepackage{subfigure}

\ifCLASSINFOpdf
\else
\fi
\begin{document}
%


\title{Data offloading in mobile edge computing: A coalitional game based pricing approach}

\author{Tian Zhang,
      Wei Chen,~\IEEEmembership{Senior Member,~IEEE,} and Feng Yang
\thanks{T. Zhang and F. Yang are with the School of Information Science and Engineering, Shandong Normal University, Jinan 250014,
China, e-mail: tianzhang.ee@gmail.com.}
\thanks{W. Chen is with Tsinghua National Laboratory for Information Science and
 Technology (TNList), Department of Electronic Engineering, Tsinghua Universtiy, Beijing 100084, China, e-mail: wchen@tsinghua.edu.cn.}
}


%


\maketitle

\begin{abstract}
Mobile edge computing (MEC), affords service to the vicinity of mobile devices (MDs), has become a key technology for future network. Offloading big data to the MEC server for preprocessing is a attractive choice of MDs.
In the paper, we investigate data offloading from MDs to MEC servers. A coalitional game based pricing scheme is proposed. We apply coalitional game to depict the offloading relationship between MDs and MEC servers, and utilize pricing as the stimuli for the offloading. A scheduled MD chooses one MEC server within the same coalition for offloading, and pays the selected MEC server for the MEC service.
We formulate a coalitional game, where MDs and MEC servers are players and their utilities are respectively defined. Next, we analyze the formulated game. Specially, the core is studied. Finally, utility performance of the proposed scheme under the 2-MD and 2-MEC- server scenario are demonstrated.
\end{abstract}


%
\IEEEpeerreviewmaketitle

\section{Introduction}
Mobile edge computing (MEC) enabling low-latency, high-bandwidth, and agile mobile services has attracted much attention in both academia and industry \cite{Whitepaper2014: M. Patel}-\cite{IEEEIoT: N. Abbas Y. Zhang A. Taherkordi T. Skeie}. MEC is a paradigm that provides cloud services within the vicinity of mobile device via the radio access network. In contrast to the service centralization of mobile cloud computing (MCC), MEC aims to empower the network edge. Since the proximity, MEC has the virtue of low latency, mobile energy saving, privacy and security enhancement, content-awareness.
Especially, one advantage is pre-processing of large data at the MEC server before sending it to the cloud.
As fusion of wireless communications and mobile computing, MEC is viewed as a key technology of next generation networks, e.g., 5G, Internet of Things (IoT), Internet of Me, Tactile Internet, Social Networks, etc.

As a novel and hot research area, there are some works on MEC. In \cite{IEEETCOM17:T. Q. Dinh  J. Tang  Q. D. La  and T. Q. S. Quek}, the offloading from a single mobile device (MD) to multiple edge devices has been investigated under a proposed optimization framework. And the authors focus on the task allocation and computational frequency scaling. In \cite{IEEETVT17:C. Wang  F. R. Yu C. Liang  Q. Chen and L. Tang}, computation offloading together with interference management have been jointly studied under an integrated framework. In \cite{IEEEWCL17:X. Tao  K. Ota  M. Dong  H. Qi  and K. Li}, performance guaranteed computation offloading has been investigated for mobile-edge cloud computing. An energy minimizing optimization problem is proposed and solved by apply Karush-Kuhn-Tucker (KKT) conditions. In \cite{IEEEICC17: T. Zhao S. Zhou  X. Guo  and Z. Niu}, resource allocation and tasks scheduling in heterogeneous cloud have been studied for delay-bounded MEC. Maximization for the probability of meeting the delay-requirements is proved to be concave and an optimal algorithm is proposed. In \cite{IEEEACCESS17: H. Zhang F. Guo H. Ji and C. Zhu}, the authors have investigated matching problem between the MEC service providers (SPs) and the user equipments (UEs) under a multi-MEC and multi-UE scenario. In \cite{IEEEICC17: F. Wang J. Xu  X. Wang and S. Cui}, joint offloading and computing optimization have been investigated in wireless powered MEC systems. Energy cost minimization of access point is formulated in convex framework and optimal solution in semi-closed form is derived. In \cite{IEEEICDCS17: Y. Li Y. Chen  T. Lan and G. Venkataramani}, the tradeoff between service response time and Quality-of-Result (QoR) has been identified in MEC. A new optimization framework that minimizes app energy consumption and service response time is proposed by jointly optimizing the QoR and the offloading strategy. In \cite{IEEEIoT: A. Kiani and N. Ansari}, an auction-based profit maximization method has been proposed towards the hierarchical MEC. In \cite{IEEETCCN: J. Xu L. Chen and S. Ren}, energy harvesting aided MEC has been investigated, and an online learning algorithm for offloading and auto-scaling is proposed.

Coalitional game theory has been widely applied in design and analysis of wireless communications system especially when cooperation occurs \cite{IEEETSM09: W. Saad Z. Han M. Debbah A. Hjorungnes and Tamer Basar}. In \cite{IEEETBDATA: Z. Su and Q. Xu}, a security-aware resource allocation approach of delivering mobile social big data has been proposed by utilizing joint match-coalitional game. In \cite{IEEETSC: O. A. Wahab J. Bentahar H. Otrok and A. Mourad}, a trust-based hedonic coalitional game has been formulated towards trustworthy multi-cloud services communities, and a three-fold solution is derived thereafter. In \cite{IEEEWCNC15: X. Lu P. Wang and D. Niyato}, A layered coalitional game algorithm is given for hierarchical cooperation in operator-controlled device-to-device (D2D) communications. In \cite{IEEEICC13}, by using coalitional game theory, we have studied cooperative transmission in vehicular networks.

Traditional data collection of edge MDs are transferred to the core network directly with high bandwidth cost and latency.
Instead, data collection is performed at the edge of the network in MEC. Bandwidth consumption is greatly reduced and latency performance improves. We investigate the data offloading of MEC in the paper. Specifically, data of MDs are transmitted to neighboring MEC servers (offloading from MDs to MEC servers), where preprocessing (e.g., data analytic) is performed, over fading channels. After data preprocessing, the results are sent to the cloud server (over wire channels, e.g., fiber channels). As multiple MDs and multiple MEC servers are considered, the data offloading from MDs to MEC server becomes a challenge problem. \emph{How to schedule multiple MDs' offloading?}\emph{How to choose the MEC servers for offloading?} \emph{How to encourage offloading among MDs and MEC servers so as to attain the virtue of MEC?}

 Utilizing coalitional game theory and pricing mechanism, we propose a joint coalition-pricing based data offloading approach. First, MDs and MEC servers form different coalitions. In each coalition, MDs are scheduled for data offloading to selected MEC server within the same coalition. Selection of MEC servers is closely related to our proposed notion \lq\lq coverage probability \rq\rq~ which shows the availability of MEC servers to MDs. Furthermore, pricing mechanism is utilized to promote cooperation between MDs and MEC servers. In one hand, MDs could gain throughput increase when utilizing MEC server (compared with no MEC). On the other hand, MEC server charges the MDs that choose it as MEC service provider in each data offloading.

The rest of paper is organized as follows. The considered system model is described in Section \ref{SectionSystem-model}. In Section \ref{SectionGame-Formulation}, a coalitional game is formulated. Next, the game analysis is preformed in Section \ref{SectionSystem-Analysis}. Numerical results are illustrated in Section \ref{SectionSimulations}. Finally, Section \ref{SectionConclution} concludes the whole paper.

\section{System model}\label{SectionSystem-model}

\begin{figure}[!t]
\centering
\includegraphics[width=3.6in]{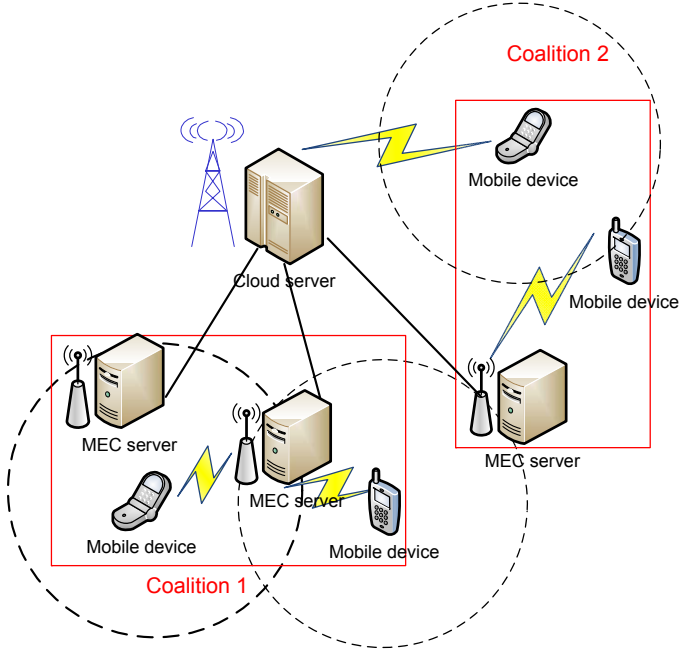}
\caption{MEC network.}
\label{Fig-SysModel}
\end{figure}
As illustrated in Fig. \ref{Fig-SysModel}, consider an MEC network consisting of a cloud server, $L$ MEC servers, and $K$ MDs.
The cloud server is represented as $0$. $\mathbb{S}=\{1,\cdots,L\}$ and $\mathbb{M}=\{L+1,\cdots,L+K\}$ denote the sets of MEC servers and MDs, respectively. Denote the transmit distance of MD $i\in\mathbb{M}$ as $d_i$. That is to say, within the sphere with radius $d_i$ around MD $i$ (referred to as transmit range), an MEC server could correctly receive the transmitted data. When MEC server $j\in \mathbb{S}$ is in the transmit range of MD $i$, we call \lq\lq $j$ covers $i$ \rq\rq. Mobile device $i\in \mathbb{M}$ is active with probability $p_i$, independently of other MDs, in each slot. When two or more mobile devices transmit simultaneously, it is referred to as a collision. We assume that the transmissions fail once a collision occurs.

MEC servers and MDs form coalitions to facilitate data transmission. Mobile devices in the same coalition can cooperate to schedule the data transmission, and only one MD can transmit at a slot to avoid interference. MEC servers that are in the same coalition with MD $i$ and cover MD $i$ constitute the feasible MEC server set of MD $i$.
In each coalition, a scheduler chooses an active mobile that can transmit while other active MDs remain silent.

When MD $i$ is scheduled to transmit at a time-slot, it selects one MEC server from its feasible MEC server set, and sends its data to the the selected MEC server for preprocessing and forwarding. If the feasible MEC server set is empty, i.e., no MEC server is available, the MD transmits to the cloud server directly (with a lower rate given fixed power). Then one MD only transmits to at most one MEC server for
each transmission. Furthermore, as no more than one MD
is scheduled to transmit at a given time-slot, one MEC server communicates with
at the most one MD during a given time-slot. MEC server acts as a helper in data transmission from the MD to the cloud server. When MD $i$ utilizes MEC server $j$ for its transmission, MEC server $j$ charges
MD $i$ with price $\xi_{ji}$ per transmission.

When some MDs and MEC servers form a coalition, the MDs
would share the channel with each other in time division multiple access (TDMA)
mode, and they need to pay for the MEC servers. However,
the MDs can avoid collision and improve bandwidth efficiency and latency. On the other hand, although there are costs in
receiving, processing and forwarding the MDs¡¯ signals, the MEC servers
could achieve revenues by charging the MDs. In a word,
both MDs and MEC servers have incentives to form coalitions.

\section{Coalitional game formulation}\label{SectionGame-Formulation}

A coalitional game $\mathcal{G}$ is uniquely defined by the pair $(\mathcal{N},v)$, where $\mathcal{N}$ is the set of players, and $v$ is the coalition value quantifying the worth of a coalition in a game. Any non-empty subset of $\mathcal{N}$ is referred to as a coalition. In the paper, players are the MDs and the MEC server. That is to say, $\mathcal{N}=\mathbb{S}\cup \mathbb{M}$. $S\subseteq \mathcal{N}$ is a coalition. Denote $S_s=S\cap \mathbb{S}$ and $S_m=S\cap \mathbb{M}$. Consider a slot, let $i\leftrightarrow j$ denote MEC server $j$ covers MD $i$ during the whole slot and $\mathcal{P}_{ji}=\Pr\{ i\leftrightarrow j\}$.

The per bandwidth data rate increase of MD $i$ with the edge computing of the MEC server
$j$ is denoted as $\Delta_{ij}$.\footnote{Since the bandwidth efficiency and latency improves with the edge computing, then the equivalent per bandwidth data rate from the MD to the cloud server increases.} Then the per bandwidth rate of MD $i$ when utilizing MEC server $j$ can be expressed as $R_{ij}=r_i+\Delta_{ij}$ with $r_i$ being direct transmission rate to the cloud server under the same power.\footnote{Specially, when direct transmission is impossible, $r_i=0$}

Formally, the scheduler in $S$ is a map $f_S: 2^{S} \to S$ such that $f_S(\Psi) \in \Psi$ for all $\Psi \subseteq S$ and $f_S(\Psi)=\emptyset$ iff $\Psi=\emptyset$.

The average effective throughput for MD $i$ can be expressed as $$T_i(S)=\mathbb{E}_{\Psi}\left\{\textbf{1}_{f_S(\Psi)=\{i\}}\right\}(\epsilon_i(S)+\zeta_i(S))\prod\limits_{j\in \mathbb{M}\backslash S_m}\left(1-p_j\right),$$
where
$\mathbb{E}_{\Psi}\left\{\textbf{1}_{f_S(\Psi)=\{i\}}\right\}$ denotes the ratio of time-slots that MD $i$ is chosen to transmit. For example, we can assume $f_S(\Psi)$ chooses the minimal element from the set of feasible transmitting MD $\Psi$.
Denote $S_m=\left\{s_1,\cdots,s_{|S_m|}\right\}$ with $s_1>\cdots >s_{|S_m|}$, then
\begin{eqnarray}\label{ratio for the minimal elemetn in Su}
\mathbb{E}_{\Psi}\left\{\textbf{1}_{f_S(\Psi)=\{s_{|S_m|}\}}\right\}=p_{s_{|S_m|}}
\end{eqnarray}
 and
 \begin{eqnarray}\label{ratio for k-th lagrest the elemetn in Su}
 \lefteqn{
 \mathbb{E}_{\Psi}\left\{\textbf{1}_{f_S(\Psi)
=\{s_{k}\}}\right\}
 }
 \nonumber \\
&=&p_{s_{k}}\prod\limits_{i=k+1}^{|S_m|}(1-p_{s_i}),k=1,\cdots,|S_m|-1.
  \end{eqnarray}
In the paper, we assume that selection of the MEC server is performed according to a uniform probability distribution for simplicity.
The average data rate for MD $i$ with mobile edge computing, $\zeta_i(S)$, is given by
\begin{eqnarray}
\lefteqn{
\zeta_i(S)=\sum\limits_{j\in S_s}R_{ij}\mathcal{P}_{ji}\prod\limits_{k \ne j\in S_s}(1-\mathcal{P}_{ki})
}
\nonumber \\
&+&\sum\limits_{j<k  \in S_s}\frac{R_{ij}+R_{ik}}{2}\mathcal{P}_{ji}\mathcal{P}_{ki}\prod\limits_{l \ne j,l \ne k \in S_s}(1-\mathcal{P}_{li})\nonumber \\
&+&\sum\limits_{j<k <l\in S_s}\frac{R_{ij}+R_{ik}+R_{il}}{3}\mathcal{P}_{ji}\mathcal{P}_{ki}\mathcal{P}_{li}\nonumber \\
&\times&\prod\limits_{r \ne j,r \ne k,r\ne l \in S_s}(1-\mathcal{P}_{ri})
+
\cdots
+\frac{\sum\limits_{j\in S_s}R_{ij}}{n}\prod\limits_{j\in S_s}\mathcal{P}_{ji}, \nonumber
\end{eqnarray}
where $n=|S_s|$. If $S_s=\emptyset$, i.e., $n=0$, $\zeta_i(S)=0$. The average direct rate
$$\epsilon_i(S)=r_i*(1-\mathcal{P}_i),$$ where $$\mathcal{P}_{i}=1-\prod\limits_{j\in S_s}\left(1-\mathcal{P}_{ji}\right)$$ is the probability that at least one MEC server in the coalition covers MD $i$.
Suppose $\Delta_{ij}=\Delta_{i}$, i.e., the data increase for MD $i$ is irrelevant to the selection of MEC servers, then $R_{ij}=\mathcal{R}_i$ and
the average throughput for MD $i$ can be simplified as
$$
T_i(S)=\mathbb{E}_{\Psi}\left\{\textbf{1}_{f_S(\Psi)=\{i\}}\right\}(\mathcal{P}_{i}\mathcal{R}_i+r_i*(1-\mathcal{P}_i))\prod\limits_{j\in \mathbb{M}\backslash S_m}\left(1-p_j\right)
$$

\emph{Remark: If there are active MDs outside $S$ at a given time-slot, at least one MDs outside $S$ transmits simultaneously with the scheduled MD in $S$ no matter how the MDs outside $S$ form coalitions. Thus, there is no collision if and only if (iff) all MDs outside $S$ is inactive. }

For MD $i$, the average payment made to the MEC servers can be given by
$$
P_i(S)=\mathbb{E}_{\Psi}\left\{\textbf{1}_{f_S(\Psi)=\{i\}}\right\}\chi_i(S)
$$
with
\begin{eqnarray}
\lefteqn{
\chi_i(S)=
\sum\limits_{j\in S_s}\xi_{ji}\mathcal{P}_{ji}\prod\limits_{k \ne j\in S_s}(1-\mathcal{P}_{ki})
}
\nonumber \\
&+&\sum\limits_{j<k  \in S_s}\frac{\xi_{ji}+\xi_{ki}}{2}\mathcal{P}_{ji}\mathcal{P}_{ki}\prod\limits_{l \ne j,l \ne k \in S_s}(1-\mathcal{P}_{li})\nonumber \\
&+&\sum\limits_{j<k <l\in S_s}\frac{\xi_{ji}+\xi_{ki}+\xi_{li}}{3}\mathcal{P}_{ji}\mathcal{P}_{ki}\mathcal{P}_{li}\nonumber \\
&\times&\prod\limits_{r \ne j,r \ne k,r\ne l \in S_s}(1-\mathcal{P}_{ri})
+
\cdots
+\frac{\sum\limits_{j\in S_s}\xi_{ji}}{n}\prod\limits_{j\in S_s}\mathcal{P}_{ji}.\nonumber
\end{eqnarray}
If $n=0$, $\chi_i(S)=0$.
Assume $\xi_{ji}=\xi_i$, i.e., all the MEC servers set the same price for MD $i$, then
the average payment can be simplified as
$$
P_i(S)=\mathbb{E}_{\Psi}\left\{\textbf{1}_{f_S(\Psi)=\{i\}}\right\}\mathcal{P}_{i}\xi_i.
$$
\par
\emph{Remark: The MEC server charges the MD once the MDs employs the MEC server for a transmission, it does not take the collisions into account. That is to say, the actions in other coalitions do not affect the charging.}
\par
The payoff of MD $i$ is determined by
$$
u_i(S)=\alpha_i T_i(S)-\beta_iP_i(S).
$$

\par
For MEC server $j$ in $S$, the revenue charged from the MDs can be given by
\begin{eqnarray}\label{R}
R_j(S)=\sum\limits_{i\in S_m}\mathbb{E}_{\Psi}\left\{\textbf{1}_{f_S(\Psi)=\{i\}}\right\}\eta_{ij}(S)\xi_{ji},
\end{eqnarray}
where $\eta_{ij}(S)$ is the probability that MD $i$ employs MEC server $j$ for transmission, and it is given by
\begin{eqnarray}
\eta_{ij}(S)&=&\mathcal{P}_{ji}
\bigg[\prod\limits_{k \ne j\in S_s}(1-\mathcal{P}_{ki})\nonumber \\
&+&\frac{1}{2}\sum\limits_{k \ne j\in S_s}\mathcal{P}_{ki}\prod\limits_{l \ne j,l \ne k \in S_r}(1-\mathcal{P}_{li})\nonumber \\
&+&\frac{1}{3}\sum\limits_{k <l, k\ne j, l\ne j\in S_s}\mathcal{P}_{ki}\mathcal{P}_{li}\prod\limits_{r \ne j,r \ne k,r\ne l \in S_s}(1-\mathcal{P}_{ri})\nonumber \\
&+&
\cdots
+\frac{1}{n}\prod\limits_{k \ne j\in S_s}\mathcal{P}_{ki}\bigg].
\end{eqnarray}
\par
Assume that MEC server $j$ receives the signal of MD $i$ at cost $c_{ji}^{r}$ and the cost of pre-processing signal and forwarding to the cloud server as $c_{ji}^{f}$.
The average cost of MEC server $j$ can be expressed as
\begin{eqnarray}\label{C}
C_j(S)=\sum\limits_{i\in S_m}\mathbb{E}_{\Psi}\left\{\textbf{1}_{f_S(\Psi)=\{i\}}\right\}\Big[c_{ji}^{f}\eta_{ij}(S)+\mathcal{P}_{ji}c_{ji}^{r}\Big].
\end{eqnarray}
\emph{Remark: MEC server $j$ receives the message of MD $i$ once it covers MD $i$ (with probability $\mathcal{P}_{ji}$), and it pre-processes the message and forwards only when it is selected as the helper by MD $i$ (with probability $\eta_{ij}(S)$).}\par
The payoff of MEC server $j$ is determined by
$$\tilde{u}_j(S)=\gamma_j R_j(S)-\mu_jC_j(S).$$

Define $v(S) \subseteq \mathfrak{R}^{|S_s|+|S_m|}$ be the set of feasible payoff vectors for $S$,
we formulate the considered data collection problem as a coalitional NTU-game $\mathcal{G}:(\mathbb{M}\cup \mathbb{S},v)$.

\section{Coalitional game analysis}\label{SectionSystem-Analysis}

In this section, we carry out analytical analysis on the formulated game. First, we present two analytical results.
Next, we investigate the stability of the game and propose a sufficient condition for the existence of the core.
\theoremstyle{definition} \newtheorem{lemma}{Lemma}
\par
In the beginning, we have the following lemma.
\begin{lemma}\label{Cancellation}
Let
$$
f(S)=\sum\limits_{i \in S_m} u_i(S)+\sum\limits_{j \in S_s} \tilde{u}_j(S)
$$
denote the sum payoff of $S$.
When $\gamma_j=1$ and $\beta_i=1$, we have $$f(S)=\sum\limits_{i \in S_m} \alpha_iT_i(S)-\sum\limits_{j \in S_s} \mu_jC_j(S).$$ That is to say, the pricing has no effect on the sum payoff in this case.
\end{lemma}
\begin{IEEEproof}
First we can prove that
$
\chi_i=\sum\limits_{j \in S_s}\eta_{ij}(S)\xi_{ji}.
$
Then
\begin{eqnarray}\label{1}
\lefteqn{
\mathbb{E}_{\Psi}\left\{\textbf{1}_{f_S(\Psi)=\{i\}}\right\}\chi_i
}
\nonumber \\
&=&
\mathbb{E}_{\Psi}\left\{\textbf{1}_{f_S(\Psi)=\{i\}}\right\}\sum\limits_{j \in S_s}\eta_{ij}(S)\xi_{ji}
\nonumber \\
&\stackrel{(a)}{=}& \sum\limits_{j \in S_s}\mathbb{E}_{\Psi}\left\{\textbf{1}_{f_S(\Psi)=\{i\}}\right\}\eta_{ij}(S)\xi_{ji}.
\end{eqnarray}
($a$) holds since $\mathbb{E}_{\Psi}\left\{\textbf{1}_{f_S(\Psi)=\{i\}}\right\}$ is irrelevant to $j \in S_s$.
Next, based on (\ref{1}), we can derive
\begin{eqnarray}
\lefteqn{
\sum\limits_{i\in S_m}\mathbb{E}_{\Psi}\left\{\textbf{1}_{f_S(\Psi)=\{i\}}\right\}\chi_i
}
\nonumber \\
&= & \sum\limits_{i\in S_m}\sum\limits_{j \in S_s}\mathbb{E}_{\Psi}\left\{\textbf{1}_{f_S(\Psi)=\{i\}}\right\}\eta_{ij}(S)\xi_{ji}.
\end{eqnarray}
Exchanging the summation order on the right side, we get
\begin{eqnarray}\label{preProof}
\sum\limits_{i \in S_u} P_i(S)=\sum\limits_{j \in S_s} R_j(S).
\end{eqnarray}
When $\gamma_j=1$ and $\beta_i=1$,
$$
f(S)=\sum\limits_{i \in S_m} \alpha_iT_i(S)-\sum\limits_{j \in S_s} \mu_jC_j(S)
+\sum\limits_{i \in S_m} P_i(S)-\sum\limits_{j \in S_s} R_j(S).
$$
Using (\ref{preProof}), we derive $$f(S)=\sum\limits_{i \in S_m} \alpha_iT_i(S)-\sum\limits_{j \in S_s} \mu_jC_j(S).$$
\end{IEEEproof}
\emph{Remark: Lemma \ref{Cancellation} reveals the fact that the total revenues obtained by the MEC servers equal to the payments of all the MDs. }
\par
\par
In addition, we obtain the second lemma.
\begin{lemma}\label{RNnocoalition}
A coalition $S$ should have at least one MD. Otherwise, $\tilde{u}_i(S)=0=\tilde{u}_i(\{i\})$ and $v(S)=\sum\limits_{i \in S}\tilde{u}_i(S)=0$. That is to say, when there are only the MEC servers, the MEC servers have no stimuli to form coalitions and each MEC server will act alone.
\end{lemma}
\begin{IEEEproof}
When $|S_m|=0$, we get $R_j(S)=0$ and $C_j(S)=0$ according to (\ref{R}) and (\ref{C}), respectively. Thus, $\tilde{u}_i(S)=0$. Specifically, $\tilde{u}_i(\{i\})=0$. As $\tilde{u}_i(S)=\tilde{u}_i(\{i\})$ in this case, each MEC server will act alone.
\end{IEEEproof}
\emph{Remark: The function of the MEC server is pre-processing and forwarding the MD's signal. So when there is no MD, it is meaningless to group only the MEC servers together.}
\par
On the other hand,
when there is no MEC server in a coalition $S$, i.e., $S \subseteq \mathbb{M}$, $$u_i(S)=\mathbb{E}_{\Psi}\left\{\textbf{1}_{f_S(\Psi)=\{i\}}\right\}*r_i*\prod\limits_{j\in \mathbb{M}\backslash S}\left(1-p_j\right)$$ for $i \in S$ and
$v(S)=\sum\limits_{i \in S}u_i(S)>0$. Specially when $S=\{i\}$, we derive $$u_i(\{i\})=p_ir_i\prod\limits_{j\in \mathbb{M} \backslash \{i\}}\left(1-p_j\right).$$ Hence, when $\exists S \subseteq \mathbb{M}~\&~ S \owns i$ satisfying $u_i(S)>u_i(\{i\})$, the MDs will form coalitions to improve the utility. Specifically, let $S=\{s_1,\cdots,s_{|S|}\}$ with $s_1>\cdots>s_{|S|}$, based on (\ref{ratio for the minimal elemetn in Su}) and (\ref{ratio for k-th lagrest the elemetn in Su}), we can derive that if
\begin{eqnarray}\label{condition for forming coalition for pure MD}
1 \ge \left\{
\begin{array}{ll}
\prod\limits_{j \in S\backslash \{s_i\}}\left(1-p_j\right), & i=|S|;\\
\frac{\prod\limits_{j \in S\backslash \{s_i\}}\left(1-p_j\right)}
{\prod\limits_{k =i+1}^{|S|}\left(1-p_{s_k}\right)} , &  \mathrm{otherwise},
\end{array} \right.
\end{eqnarray}
forming coalition $S$ is profitable.\footnote{Although forming $S$ may be not optimal, it is at least better than acting alone.}
Specially, when $p_i=p$, i.e., all MDs have the same active probability, we can derive that (\ref{condition for forming coalition for pure MD}) holds,
then forming coalitions is always profitable in the case. The reason is as follows: Since the formulation of coalitions can avoid collision,
then it can improve the effective throughput.

\theoremstyle{definition} \newtheorem{definition}{Definition}
\par
The definition for the core of our coalitional game is given as follows.
\begin{definition}
The core of $(\mathbb{M}\cup \mathbb{S},v)$ is defined as $$C=\big\{x\in  v\left(\mathbb{M}\cup \mathbb{S}\right): \forall S, \not\exists y \in v(S), s.t. ~ y_i>x_i, \forall i \in S\big\}.$$
\end{definition}
\par
The following lemma gives a sufficient condition for the existence of the core.
\begin{lemma}\label{sufficient condition for the existence of the core}
The core of $(\mathbb{M}\cup \mathbb{S},v)$ is nonempty once the following conditions hold ($S \subset \mathbb{M}\cup \mathbb{S}$):
\begin{enumerate}[1)]
\item $\alpha_i>0$, $\beta_i>0$, $\gamma_j>0$, and $\mu_j>0$.
\item  $\alpha_i T_i(S)>\beta_iP_i(S)$ or $ \gamma_j R_j(S)>\mu_jC_j(S)$.
\item $ \alpha_i T_i(\mathbb{M}\cup \mathbb{S})-\beta_iP_i(\mathbb{M}\cup \mathbb{S})>\alpha_i T_i(S)-\beta_iP_i(S)$, and $\gamma_j R_j(\mathbb{M}\cup \mathbb{S})-\mu_jC_j(\mathbb{M}\cup \mathbb{S})>\gamma_j R_j(S)-\mu_jC_j(S)$.
\end{enumerate}
\end{lemma}
\begin{IEEEproof}
When 1) holds, we can find $\alpha_i$, $\beta_i$, $\gamma_i$, and $\mu_j$ to satisfy 2).  If 2) does not holds,
we have
$$u_i(\{i\})=p_ir_i\prod\limits_{j\in \mathbb{M}/\{i\}}\left(1-p_j\right)\ge 0\ge u_i(S)$$ for $i \in S_m$
and $\tilde{u}_j(S)\le 0=\tilde{u}_j(\{j\})$ for $j \in S_s$. Then, each MD and MEC server will act alone. In this case, the core is empty. When 3) holds, we can prove that $(\mathbb{M}\cup \mathbb{S},v)$ is balanced \cite{Book91:R. B. Myerson}. Thus, the core is nonempty according to the Bondareva-Shapley theorem.
\end{IEEEproof}

\section{Numerical results}\label{SectionSimulations}

In the simulations, we consider 2 MDs (MD 1 and MD 2) and 2 MEC servers (MEC-s 3 and MEC-s 4). Two dimensional scenario is considered. MDs are uniformly distributed
in an 2-dimensional area with $10\times 10$, where $(5,5)$ is its center, MEC-s 3 locates at $(0,0)$, MEC-s 4 locates at $(7,6)$.
$10^6$ topology implementations are averaged to get the mean performance.

\begin{figure}[!t]
\centering
\includegraphics[width=3.6in]{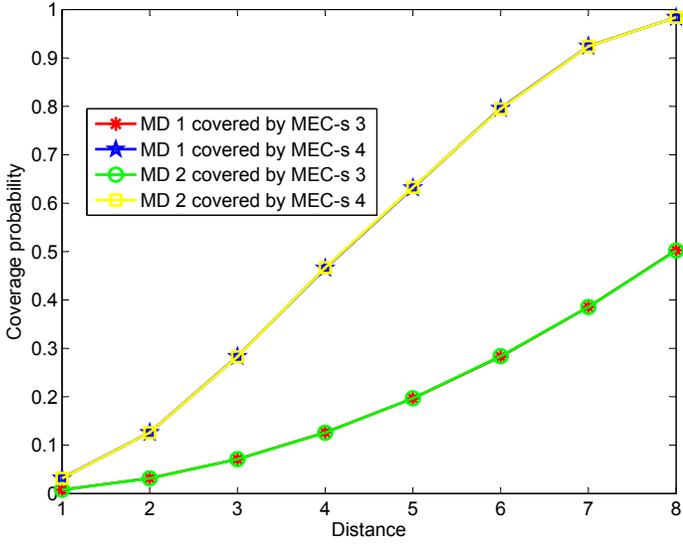}
\caption{Coverage probability v.s. transmit distance, $d$}
\label{Sim1-Coverage}
\end{figure}

\begin{figure}[!t]
\centering
\includegraphics[width=3.6in]{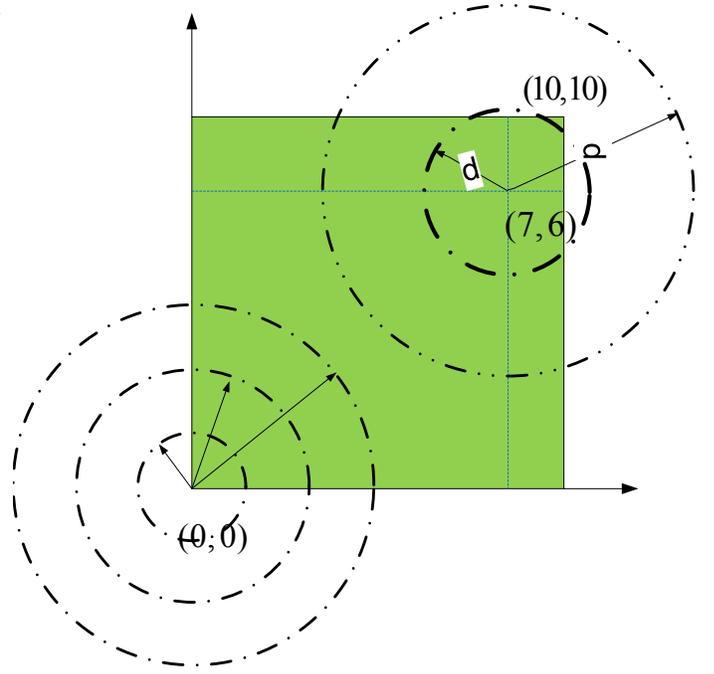}
\caption{Theoretical computation of the coverage probability when MDs are uniformly distributed}
\label{Sim1-CoverageExplaination}
\end{figure}

Fig. \ref{Sim1-Coverage} shows the probability of MD 1 and MD 2 being covered by MEC-s 3 and MEC-s 4, respectively. From the figure, we can find that the coverage probabilities of MEC-s 3 over MD 1 and MD 2 are same, and the coverage probabilities of MEC-s 4 over MD 1 and MD 2 are same also. In addition, the coverage probability of MEC-s 4 is higher than that of MEC-s 3 and both increase with the increase of $d$. We can explain by theoretical analysis. Since MD 1 and MD 2 are uniformly distributed, the coverage probability can be computed using Fig. \ref{Sim1-CoverageExplaination} in two-dimensional case. For any MD, when it is in the overlap of a circle centering at $(0,0)$ with radius $d$, it can be coverage by MEC-s 3. Then, the coverage probability is the ratio between the area of the overlap and $10 \times 10$. With the increase of $d$, the ratio increases (until it reaches 1). For the MEC-s 4, the center becomes $(7,6)$ and we can obtain similar results. Furthermore, as (7,6) is nearer to $(5,5)$ (i.e., area center) with the same $d$, its overlap area is bigger than that for $(0,0)$. Then the coverage probability of MEC-s 4 is higher than that of MEC-s 3.

There are 15 possible coalitional structures for 2-MD and 2-MEC-s, as listed in Table \ref{tab:1}. Applying Lemma \ref{RNnocoalition},
$\mathcal{C}_{11}$ is similar as $\mathcal{C}_{3}$ and $\mathcal{C}_{12}$ is similar as $\mathcal{C}_{4}$.
Meanwhile, as MD 1 and MD 2 as well as MEC-s 3 and MEC-s 4 are role-exchangeable in the scheme, we can see that $\mathcal{C}_8$ is similar as $\mathcal{C}_2$; $\mathcal{C}_9$ is similar as
$\mathcal{C}_6$; $\mathcal{C}_{10}$, $\mathcal{C}_{14}$ and $\mathcal{C}_{15}$ are similar as $\mathcal{C}_5$; and $\mathcal{C}_{13}$ is similar as $\mathcal{C}_7$.
In conclusion, we need to consider $\mathcal{C}_1-\mathcal{C}_7$. In the following simulations, $\mathcal{R}_i=1.8$, $r_i=1$, $\xi_i=1.5$, $\alpha_i=10$, $p_i=0.6$, $c_{ji}^{r}=0.2$, $c_{ji}^{f}=0.5$, and $\beta_i=\gamma_i=\mu_i=1$.

\begin{table}[!t]
\caption{coalition structure for 4 nodes}
\label{tab:1}
\centering
\begin{tabular}{|l|l|l|l|}
  \hline
   $\mathcal{C}_1$: \{1,2,3,4\}& $\mathcal{C}_6$: \{1,3\},\{2,4\} &   $\mathcal{C}_{11}$: \{1,2\},\{3,4\}\\
   $\mathcal{C}_2$: \{1,3,4\},\{2\}& $\mathcal{C}_7$: \{1,2,3\},\{4\} &  $\mathcal{C}_{12}$: \{1\},\{2\},\{3,4\}  \\
   $\mathcal{C}_3$: \{1,2\},\{3\},\{4\}& $\mathcal{C}_8$: \{1\},\{2,3,4\} &  $\mathcal{C}_{13}$: \{1,2,4\},\{3\}  \\
  $\mathcal{C}_4$: \{1\},\{2\},\{3\},\{4\} & $\mathcal{C}_9$: \{1,4\},\{2,3\} &    $\mathcal{C}_{14}$: \{1,4\},\{2\},\{3\} \\
  $\mathcal{C}_5$: \{1\},\{3\},\{2,4\} & $\mathcal{C}_{10}$: \{1\},\{4\},\{2,3\}  &   $\mathcal{C}_{15}$: \{2\},\{4\},\{1,3\}\\
  \hline
\end{tabular}
\end{table}

\begin{figure}[!t]
\centering
\includegraphics[width=3.6in]{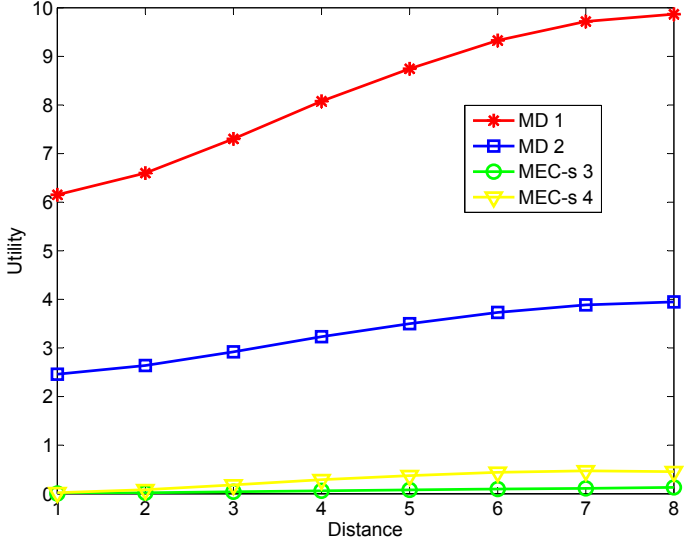}
\caption{Mean utility of $\mathcal{C}_1$}
\label{Sim-C1}
\end{figure}

Fig. \ref{Sim-C1} plots the mean utility performance under $\mathcal{C}_1$, which is also referred to as the grand coalition, i.e., all nodes form one coalition. We can observe that the utility performance improves when the transmit distance becomes longer. It is because that with the expand of the transmit distance, the coverage probability increases. Then the MEC severs could take part in the data collection with higher probability, and the rate improves. The performance of MD 1 is better than that of MD 2 since MD 1 has  priority in transmitting node selection in the scheme when MD 1 and MD 2 in the same coalition. MEC-s 4 has better coverage probability which leads to more revenues charged from MDs. Hence MEC-s 4 has better performance than MEC-s 3.

\begin{figure}[!t]
\centering
\includegraphics[width=3.6in]{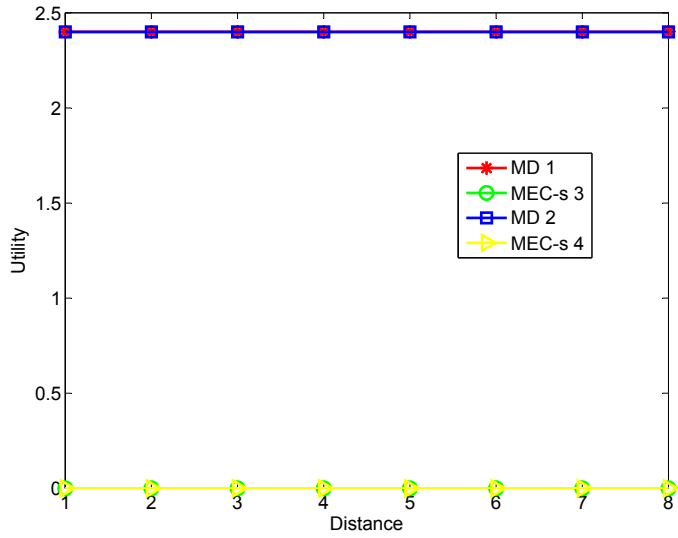}
\caption{Mean utility of $\mathcal{C}_4$}
\label{Sim-C4}
\end{figure}
Fig. \ref{Sim-C4} demonstrates the mean utility performance under $\mathcal{C}_4$, where each node forms a coalition and transmit solely. We can find that the utility performance of MEC-s 3 and MEC-s 4 remains as zero. This can be explained by using Lemma \ref{RNnocoalition}. The utility performance of MD 1 and MD 2 are the same and remain constant. The $\mathcal{C}_4$ can be viewed as a baseline.

\begin{figure}[!t]
\centering
\includegraphics[width=3.6in]{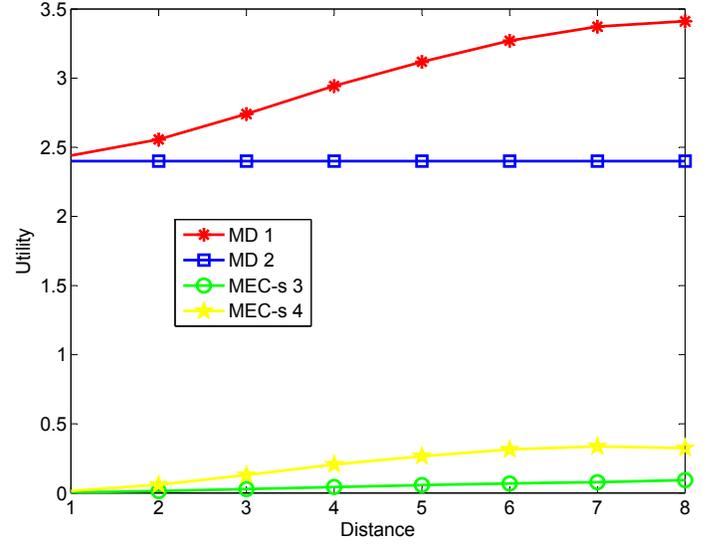}
\caption{Mean utility of $\mathcal{C}_2$}
\label{Sim-C2}
\end{figure}

Fig. \ref{Sim-C2} shows the mean utility performance under $\mathcal{C}_2$. MD 1, MEC-s 3, and MEC-s 4 are in a coalition. The performance improves with the increase of transmit distance. As MD 2 transmits solely, the utility remains constant with respect to transmit distance. The performance of MD 1 is better than that of MD 2. It verifies that the MEC could improve utility performance (especially by comparison with Fig. \ref{Sim-C4}).

\begin{figure}[!t]
\centering
\includegraphics[width=3.6in]{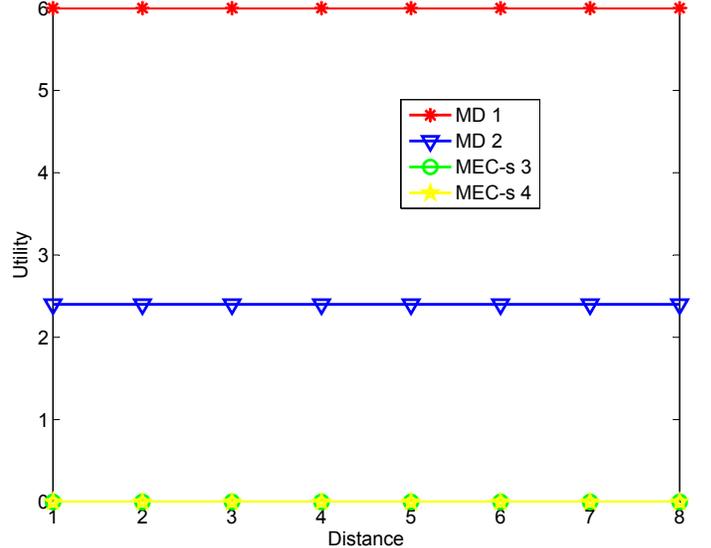}
\caption{Mean utility of $\mathcal{C}_3$}
\label{Sim-C3}
\end{figure}

Fig. \ref{Sim-C3} illustrates the utility performance with $\mathcal{C}_3$. MEC-s 3 and MEC-s 4 respectively form a coalition. And the utility performance remain zero, which complies with Lemma \ref{RNnocoalition}. MD 1 and MD 2 form one coalition. Since no MEC-s available, the utility performance are constant. The performance of MD 1 is better than that of MD 2 because of its priority in transmitting selection within the coalition. In addition, compared to Fig. \ref{Sim-C4}, we can see the utility increase of MD 1. That is to say, even without MEC server, forming coalition is profitable for MD. This can be explained as follows: When forming coalition, collisions can be avoided. Then the effective throughput increases and utility improves accordingly.

\begin{figure}[!t]
\centering
\includegraphics[width=3.6in]{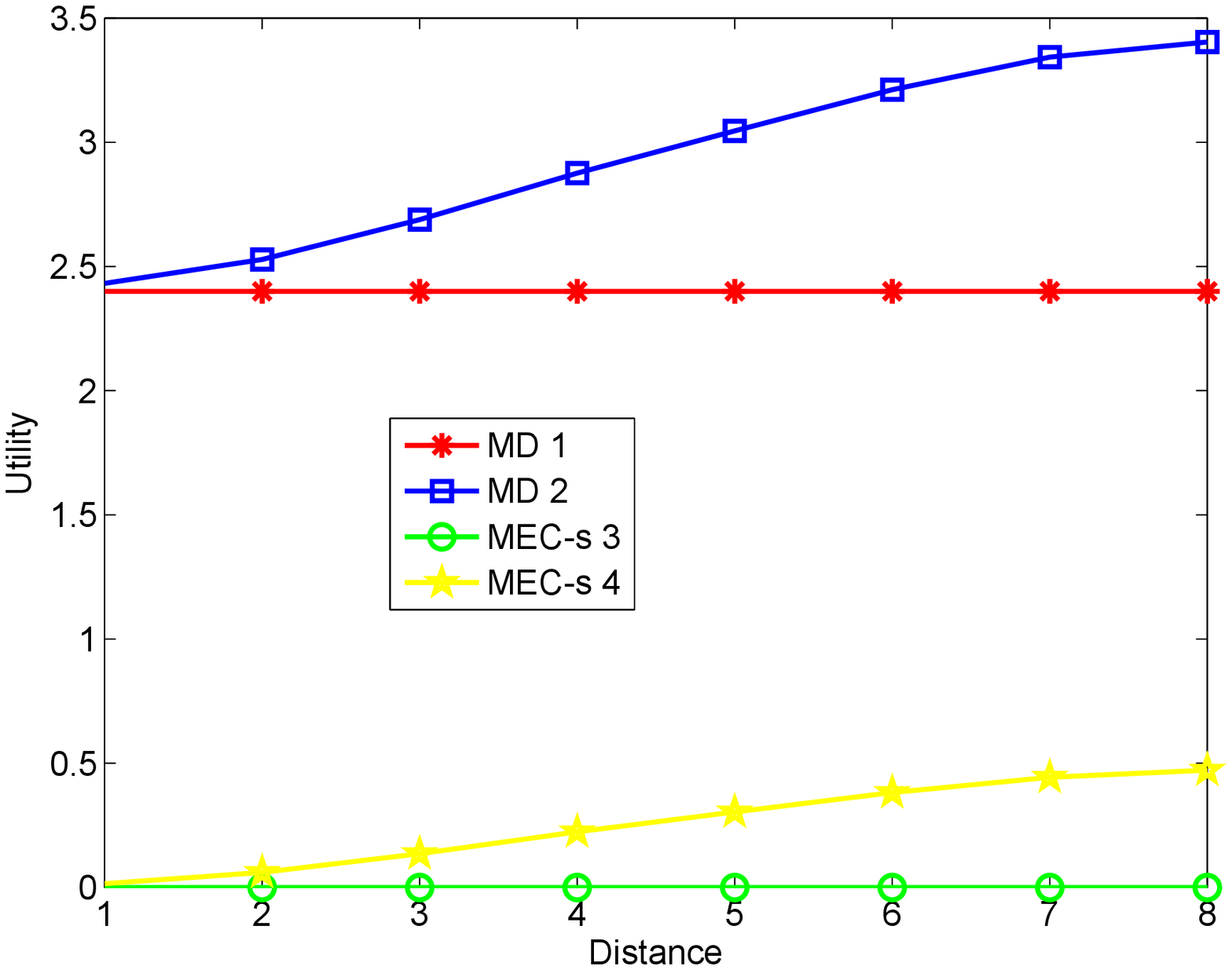}
\caption{Mean utility of $\mathcal{C}_5$}
\label{Sim-C5}
\end{figure}

Fig. \ref{Sim-C5} depicts the average utility performance under $\mathcal{C}_5$ setting. MD 1 and MEC-s 3 respectively constitute a coalition. Then the utility performance remains constant. MD 2 and MEC-s 4 form one coalition, and the utility increase with transmit distance. Furthermore, MD 2 has better performance than MD 1 and MEC-4 has better performance than MEC-s 3. This verifies the advances of constituting coalition.

\begin{figure}[!t]
\centering
\includegraphics[width=3.6in]{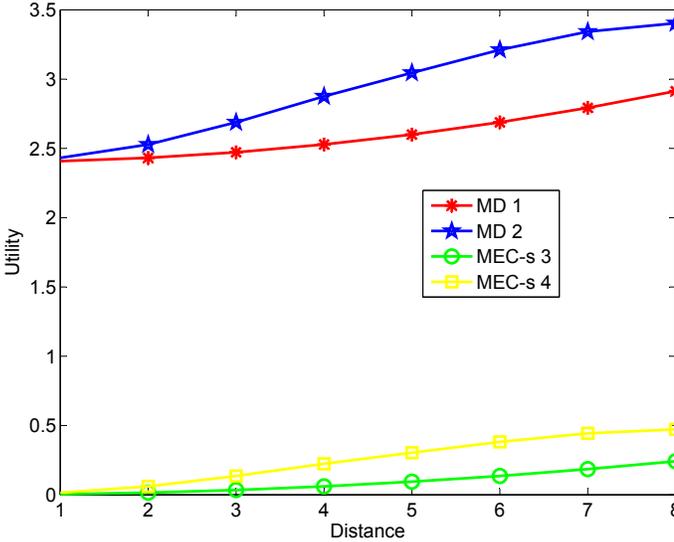}
\caption{Mean utility of $\mathcal{C}_6$}
\label{Sim-C6}
\end{figure}

Fig. \ref{Sim-C6} draws the mean utility performance with $\mathcal{C}_6$. In the setting, MD 1 and MEC-s 3 form one coalition; MD 2 and MEC-s 4 constitute another coalition. Compared with Fig. \ref{Sim-C4}, utility performance of each node has improvement. Additionally, MD 2 has better performance than MD 1 and MEC-4 has better performance than MEC-s 3. This is because that MEC-s 4 has better coverage probability than MEC-s 3, then the MEC-s 4 \& MD 2 coalition has better performance.

\begin{figure}[!t]
\centering
\includegraphics[width=3.6in]{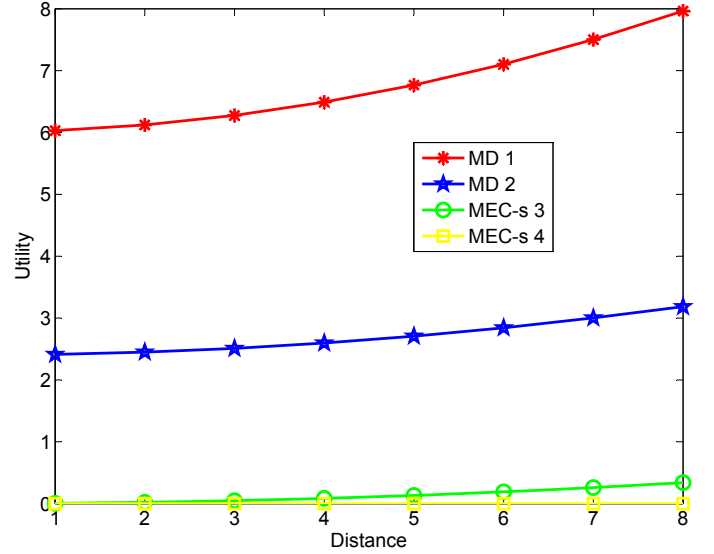}
\caption{Mean utility of $\mathcal{C}_7$}
\label{Sim-C7}
\end{figure}

Fig. \ref{Sim-C7} plots the average utility performance under $\mathcal{C}_7$. MD 1, MD 2, and MEC-s 3 constitute one coalition. MEC-s 4 forms another coalition solely. Since MEC-s 4 act alone, its utility remains zero. MD 1 and MD 2 could utilize MEC-s 3 for MEC data transmission to the cloud server. Utility performance increases with increment of $d$ since bigger value of $d$ incurs higher coverage probability and transmission rate performance thereafter. MD 1 has better performance than MD 2 as MD 1 has selection priority in scheduling.

Comparing utility performance of the considered coalition structure, we find that the grand coalition, $\mathcal{C}_1$, has the best score. In addition, all utilities in the grand coalition are positive and $\alpha_i>0$, $\beta_i=\gamma_i=\mu_i>0$. Then the conditions of Lemma \ref{sufficient condition for the existence of the core} hold, we claim that the core is not empty and $\big(u_1(\mathcal{C}_1),u_2(\mathcal{C}_1), u_3(\mathcal{C}_1), u_4(\mathcal{C}_1)\big)$ is in the core.

\section{Conclusion}\label{SectionConclution}
Data offloading in muti-MD and multi-MEC-server scenario is studied in the paper. We propose a coalition based pricing scheme. The proposed scheme incorporates MD scheduling, the MEC server selection and offloading encouragement. Theoretically, coalitional game is first formulated, and the game is analyzed thereafter. Especially, we give sufficient conditions for non-empty of the core. In the end, numerical results  verify the effectiveness of the proposed scheme by showing the utility performance.






%

\end{document}